\documentclass{iau}

\usepackage{amsmath}
\usepackage{graphicx}
\usepackage{multirow}

\def\arxivprefixesep{:}

\newcommand{\eprint}[2][]{%
	{\tt\if!#1!#2\else#1\arxivprefixesep\ignorespaces#2\fi}%
}

\newcommand{\kms}{~km\,s$^{-1}$}

\begin{document}

\lefttitle{C. Bacchini et al.}
\righttitle{Evidence for supernova-driven gas turbulence}

\jnlPage{1}{7}
\jnlDoiYr{2022}
\doival{xx.xxxx/xxxxx}

\aopheadtitle{Proceedings IAU Symposium 373}
\editors{Woong-Tae Kim and Tony Wong, eds.}

\title{Evidence for supernova feedback sustaining gas turbulence in nearby star-forming galaxies}

\author{Bacchini, C.$^{1,2,3}$; Fraternali, F.$^2$; Pezzulli, G.$^2$; Iorio, G.$^4$; Marasco, A.$^1$; Nipoti, C.$^3$}
\affiliation{
	$^1$INAF - Astronomical Observatory of Padova, Vicolo dell'Osservatorio 5, IT-35122, Padova, Italy \\email: {\tt cecilia.bacchini@inaf.it} \\
	$^2$Kapteyn Astronomical Institute, University of Groningen, Landleven 12, 9747 AD Groningen, The Netherlands \\
	$^3$Department of Physics and Astronomy, University of Bologna, via Gobetti 93/2, IT-40129, Bologna, Italy \\ 
	$^4$Department of Physics and Astronomy, University of Padova, Vicolo dell’Osservatorio 3, IT-35122, Padova, Italy \\}

\begin{abstract}
HI and CO observations indicate that the cold gas in galaxies is very turbulent. 
However, the turbulent energy is expected to be quickly dissipated, implying that some energy source is needed to explain the observations. 
The nature of such turbulence was long unclear, as even the main candidate, supernova (SN) feedback, seemed insufficient. 
Other mechanisms have been proposed, but without reaching a general consensus. 
The key novelty of our work is considering that the gas disc thickness and flaring increase the dissipation timescale of turbulence, thus reducing the energy injection rate required to sustain it. 
In excellent agreement with the theoretical expectations, we found that the fraction of the SN energy (a.k.a. SN coupling efficiency) needed to maintain the cold gas turbulence is $\sim 1$\%, solving a long-standing conundrum. 
\end{abstract}


\maketitle
\section{What drives the turbulence in the cold interstellar medium?}
In star-forming galaxies, the cold gas velocity dispersion derived from the broadening of emission lines exceeds the values expected from the gas thermal motion. 
This additional broadening is typically ascribed to turbulence. 
Turbulence in the interstellar medium (ISM) is well described by the Kolmogorov's theory \citep{1941Kolmogorov}. 
In this framework, the turbulent energy is injected into the ISM at a certain driving scale and then is transferred to the smaller scales through the so-called Kolmogorov's cascade. 
In practice, the turbulent eddies break down into smaller eddies, transferring to the kinetic energy to smaller scales. 
This energy is conserved until the dissipation scale is reached, then the turbulent energy is dissipated by the gas viscosity. 
Since we observe turbulence, this process must be in a stationary state and turbulence must be maintained by some source of new energy.  
This fresh turbulent energy must be provided with a certain injection rate depending on the dissipation timescale of turbulence. 
This latter is $\tau_\mathrm{diss} \equiv L/\upsilon_\mathrm{turb}$, with $L$ and $\upsilon_\mathrm{turb}$ being the physical scale of the system and the turbulent velocity, respectively. 
In the case of galaxies, $L$ is typically assumed proportional to the gas disc scale height $h$. 

Supernova (SN) explosions are the best candidates for being the engine of turbulence in galaxy discs, as they inject a huge amount of energy into the ISM. 
The energy flux injected by SNe can be estimated from the observed SFR ($\Sigma_\mathrm{SFR}$) using 
$\dot{E}_\mathrm{SNe} = \eta \Sigma_\mathrm{SFR} f_\mathrm{cc} E_\mathrm{SN}$, where $\eta $ is SN efficiency (or coupling efficiency), $E_\mathrm{SN}=10^{51}$~erg is the energy of a single SN, and $f_\mathrm{cc}$ is the fraction of stars exploding as SN core-collapse given an initial mass function. 
The key parameter is $\eta$, which is the fraction of the total SN energy that goes into feeding turbulence. 
Analytical and numerical models of SN remnant evolution predict $\eta \lesssim 0.1$, as most of the SN energy is radiated away \citep{1974Chevalier,2015Martizzi}. 
To understand whether SNe can sustain the cold gas turbulence in galaxies, the expected SN energy must be compared to the observed kinetic energy density of the gas $E_\mathrm{obs} = \frac{3}{2} \Sigma_\mathrm{obs} \sigma_\mathrm{obs}^2$, with $\Sigma_\mathrm{obs}$ and $\sigma_\mathrm{obs}$ being the gas surface density and velocity dispersion. 
\cite{2009Tamburro} used this approach to investigate the origin of the atomic gas turbulence in nearby galaxies, assuming $L = 100$~pc and $\upsilon_\mathrm{turb}=10$~\kms (i.e. $\tau_\mathrm{diss}=10$~Myr) across the disc. 
They found that the SN energy can explain $E_\mathrm{obs}$ in the innermost regions of their galaxies, while very high ($\eta \gg 0.1$) or even non-physical ($\eta > 1$) efficiencies were needed in the outer parts where $\Sigma_\mathrm{SFR}$ is low \citep[see also][]{2019Utomo}. 
Other mechanisms have been proposed, but no consensus has been reached about the main engine of turbulence \citep{2004MacLowKlessen}. 

\section{Slow dissipation in flaring discs}
\begin{figure}
\centering
\includegraphics[width=0.49\linewidth]{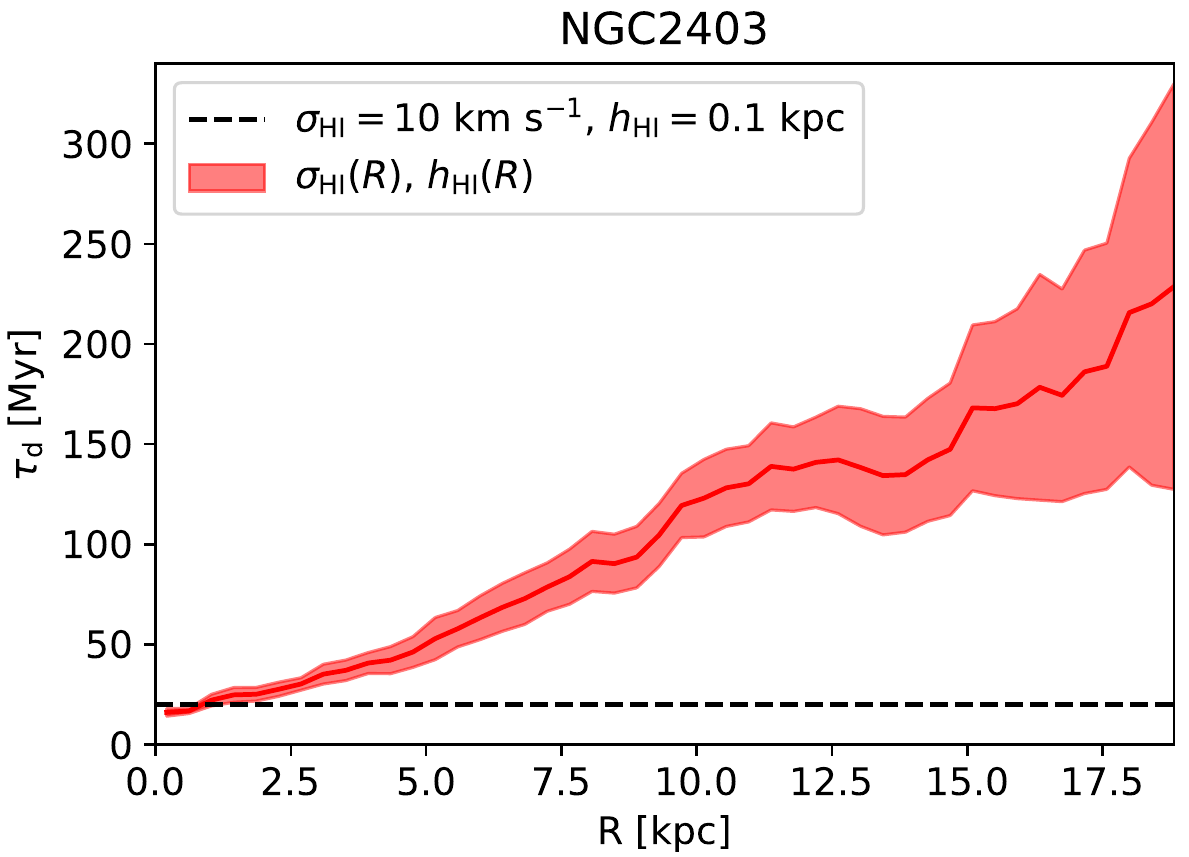}
\hspace{0.5cm}
\includegraphics[width=0.44\linewidth]{ekin.pdf}
\caption{\textit{Left:} Dissipation timescale of the atomic gas turbulence in the spiral galaxy NGC~2403. 
The dashed black line is for a thin disc with no flaring.  
The red curve is for a flaring disc in VHE (the red area shows the uncertainty). 
\textit{Right:} Observed kinetic energy density per unit area of the atomic gas in NGC~2403 (black points). 
Blue and red areas are respectively the gas thermal energy and the turbulent energy injected by SNe with the efficiency $\eta_\mathrm{atom,c}$ reported above.  
The green area is the total model energy. 
The grey dotted vertical line indicates the outermost radius with measured $\Sigma_\mathrm{SFR}$; upper limits are used for larger radii.}
\label{fig:ngc2403}
\end{figure}

The key improvement in our work \citep{2020Bacchini} consists in taking into account that the dissipation of turbulence is slower in thick and flaring discs than in thin discs without the flaring. 
Gas discs in galaxies are indeed expected to be in vertical hydrostatic equilibrium (hereafter VHE), meaning that the gas pressure, which tends to inflate the disc, balances the gravitation force towards the midplane, which tends to flatten the disc. 
Since gravity weakens with increasing distance from the galaxy centre, the gas layer thickens at large galactocentric radii (flaring). 
The natural driving scale of turbulence in galaxy discs is $L = 2h$, with $h$ being the gas disc scale height. 
As shown in the left panel of Fig.~\ref{fig:ngc2403}, the turbulence dissipation timescale in the outer regions of a flaring gas disc is one order of magnitude longer than in the case of a thin disc without the flaring, which is the case assumed by \cite{2009Tamburro}. 

We collected a sample of 10 nearby galaxies made of 2 dwarfs and 8 spirals. 
For each galaxy, we measured the distribution and kinematics of the atomic and the molecular gas from 21-cm and CO(2--1) emission line datacubes using the software  $^{\text{\textsc{3D}}}$\textsc{Barolo} \citep{2015DiTeodoro}, which performs a tilted-ring model fitting on the observations. 
Then, we calculated the radial profile of $E_\mathrm{obs}$ for both the gas phases. 
We derived $\tau_\mathrm{diss}$ using the gas scale heights based on the VHE. 
To this task, we used the Python tool \texttt{Galpynamics} \citep{2018Iorio}.
We adopted a simple model for the kinetic energy density per unit area ($E_\mathrm{mod}$) that includes the gas thermal energy and the turbulent energy expected from SNe. 
This latter depends on $\tau_\mathrm{diss}$ and the free parameter $\eta$, which we assumed to be constant across the disc. 
For the atomic gas, we explored two single-phase cases (assuming either cold or warm neutral medium), and the more realistic scenario of a two-phase gas. 
We fitted $E_\mathrm{obs}$ with $E_\mathrm{mod}$ to find $\eta$. 
The right panel in Fig.~\ref{fig:ngc2403} shows the example of the atomic gas in NGC~2403 assuming a single-phase, cold neutral gas. 
This case requires the maximum SN energy, as the gas thermal energy is essentially negligible. 
Nonetheless, $E_\mathrm{obs}$ is very well reproduced by our model, notably with very low efficiency. 
We find the same results for all the galaxies and for both the atomic gas and the molecular gas, which is remarkable considering that we assumed $\eta$ to be independent from the galactocentric radius. 
Considering the galaxy sample as a whole and the systematic uncertainties (e.g. on gas temperature and $\Sigma_\mathrm{SFR}$), we obtained $\eta = 0.02^{+0.02}_{-0.01}$ for the atomic gas and $\eta = 0.003^{+0.007}_{-0.002}$ for the molecular gas. 

\section{Concluding remarks}
The take-home message of this proceeding is that SNe can sustain the cold gas turbulence in nearby galaxies with only a few percent ($\lesssim 2$\%) of their energy. 
Hence, no additional source of energy is compulsorily required. 
The $\eta$ values obtained in this work can be useful to implement sub-grid physics in some numerical simulations of galaxy discs with limited resolution. 

In a broader context, our results can be interpreted as empirical evidence for the self-regulating cycle of star formation \citep{2010Ostriker}. 
\cite{2019Bacchini,2019Bacchini_b,2020Bacchini_b} found a tight correlation, called volumetric star formation (VSF) law, between the volume densities of the cold gas and the SFR of nearby disc galaxies. 
These volume densities were obtained from the observed surface densities using the gas scale height derived in the same way as here. 
The VHE implies that the gas scale height depends on its velocity dispersion, which includes the contributions of thermal motions and turbulence. 
The results presented here show that turbulence depends in turn on the SFR. 
Hence, we can envision the following self-regulating cycle. 
In a quasi-steady state, higher SFR means more turbulence (thicker disc) and lower gas volume density. 
According to the VSF law, the SFR declines and turbulence follows, causing the gas volume density to grow again and bringing to a new phase of high SFR.

\bibliographystyle{iaulike}
\bibliography{paty.bib}

\begin{thebibliography}{}

\bibitem[{Bacchini} et~al., 2019a]{2019Bacchini}
{Bacchini}, C., {Fraternali}, F., {Iorio}, G., \& {Pezzulli}, G. 2019,a {\em
  \aap}, 622a, A64.

\bibitem[{Bacchini} et~al., 2020a]{2020Bacchini}
{Bacchini}, C., {Fraternali}, F., {Iorio}, G., {Pezzulli}, G., {Marasco}, A.,
  \& {Nipoti}, C. 2020,a {\em \aap}, 641a, A70.

\bibitem[{Bacchini} et~al., 2020b]{2020Bacchini_b}
{Bacchini}, C., {Fraternali}, F., {Pezzulli}, G., \& {Marasco}, A. 2020,b {\em
  \aap}, 644b, A125.

\bibitem[{Bacchini} et~al., 2019b]{2019Bacchini_b}
{Bacchini}, C., {Fraternali}, F., {Pezzulli}, G., {Marasco}, A., {Iorio}, G.,
  \& {Nipoti}, C. 2019,b {\em \aap}, 632b, A127.

\bibitem[{Chevalier}, 1974]{1974Chevalier}
{Chevalier}, R.~A. 1974, {\em \apj}, 188, 501--516.

\bibitem[{Di Teodoro} and {Fraternali}, 2015]{2015DiTeodoro}
{Di Teodoro}, E.~M. \& {Fraternali}, F. 2015, {\em \mnras}, 451, 3021--3033.

\bibitem[{Iorio}, 2018]{2018Iorio}
{Iorio}, G. 2018,.
\newblock PhD Thesis, University of Bologna, (2018).

\bibitem[{Kolmogorov}, 1941]{1941Kolmogorov}
{Kolmogorov}, A. 1941, {\em Akademiia Nauk SSSR Doklady}, 30, 301--305.

\bibitem[{Mac Low} and {Klessen}, 2004]{2004MacLowKlessen}
{Mac Low}, M.-M. \& {Klessen}, R.~S. 2004, {\em Reviews of Modern Physics},
  76(1), 125--194.

\bibitem[{Martizzi} et~al., 2015]{2015Martizzi}
{Martizzi}, D., {Faucher-Gigu{\`e}re}, C.-A., \& {Quataert}, E. 2015, {\em
  \mnras}, 450(1), 504--522.

\bibitem[{Ostriker} et~al., 2010]{2010Ostriker}
{Ostriker}, E.~C., {McKee}, C.~F., \& {Leroy}, A.~K. 2010, {\em \apj}, 721,
  975--994.

\bibitem[{Tamburro} et~al., 2009]{2009Tamburro}
{Tamburro}, D., {Rix}, H.-W., {Leroy}, A.~K., {Mac Low}, M.-M., {Walter}, F.,
  {Kennicutt}, R.~C., {Brinks}, E., \& {de Blok}, W.~J.~G. 2009, {\em \aj},
  137, 4424--4435.

\bibitem[{Utomo} et~al., 2019]{2019Utomo}
{Utomo}, D., {Blitz}, L., \& {Falgarone}, E. 2019, {\em \apj}, 871(1), 17.

\end{thebibliography}

\end{document}